\documentclass[prl,manuscript,showpacs,floatfix,aps]{revtex4}
\usepackage{amssymb}
\usepackage{amsmath}
\usepackage{graphicx,bm}

\begin{document}

\title{\textbf{Fluctuating Cooper pairs in FeSe at temperatures exceeding double $T_c$}}

\author{A.\,L.\,Solovjov$^{1,2}$, E.\,V.\,Petrenko$^{1}$, L.V.\,Omelchenko$^{1}$, E.\,Nazarova$^{3}$, K.\,Buchkov$^{3}$,
and K.\,Rogacki$^{4}$}
\email {k.rogacki@intibs.pl}
\affiliation{$^1$B.\,I.\,Verkin Institute for Low Temperature Physics and Engineering, National Academy
of Science of Ukraine, 47 Nauki ave., 61103 Kharkov, Ukraine\\
$^2$Physics Department, V.\,Karazin Kharkiv National University, 4 Svobody Sq., 61077 Kharkiv, Ukraine\\
$^3$Georgi Nadjakov Institute of Solid State Physics, Bulgarian Academy of Sciences, 72 Tsarigradsko shosse Blvd., 1784 Sofia, Bulgaria\\
$^4$Institute of Low Temperature and Structure Research, Polish Academy of Sciences,
ul.\,Ok{\'{o}}lna 2, 50-422, Wroclaw, Poland}
\date{\today}

\begin{abstract}
Temperature dependencies of excess conductivity, $\sigma'$, have been studied in detail for three FeSe$_{0.94}$
textured polycrystalline samples prepared by partial melting and solid state reaction. It was revealed that both $\sigma'$ and its temperature dependence are extremely sensitive to the method of sample preparation. Then, it was shown that in the range from the superconducting transition temperature $T_c$ $\sim9$~K up to the characteristic temperature $T_{01}$ $\sim19$~K, $\sigma'(T)$ obeys the classical fluctuation theories of Aslamazov-Larkin (AL) and Hikami-Larkin (Maki-Thompson (MT) term) pointing to the existence of fluctuating Cooper pairs in FeSe at temperatures exceeding double $T_c$. Like in cuprates, AL-MT crossover at $T_0 < T_{01}$ is observed, which means the appearance of 3D-2D dimensional transition at this temperature. This allows us to determine the coherence length along the $c$-axis, $\xi_c(0)$ $\sim3$~{\AA}, and a set of additional samples' parameters, including the phase relaxation time, $\tau_{\phi}$, of fluctuating Cooper pairs, within a simple two-dimensional free-carrier picture. It was shown that $\tau_{\phi}$ in FeSe coincides with that found for YBa$_2$Cu$_3$O$_{7-\delta}$ suggesting that the nature of superconducting fluctuations is very similar for these high-temperature superconductors of different types.
\end{abstract}

\maketitle

$\bf{1.\, INTRODUCTION}$\\

Despite the fact that FeSe has the simplest crystal structure among
Fe-based superconductors, it possesses many unusual features, such as: enhanced anisotropic spin fluctuations below ~80 K \cite{Ross}, specific temperature dependences of resistivity \cite{Ross, Sun1,Kar, H2}, Hall effect \cite{Sun1,Kar} and Kohler effect \cite{Ross,Sun1,Kar}, which are likely due to presence of both electron- and hole-like charge carriers, and specific structural transition at $T_{s}\sim 90$~K \cite{ Sun1,Kar, H2,John}. Understanding these features is believed to shed more light
on the nature of high-temperature superconductors (HTSCs) in general (see Refs. \cite{Sad,Kord} and references therein), including the interplay between superconductivity and magnetism \cite{Maz,Wang,Kas,We}.
One such feature is the temperature dependence of the resistivity, $\rho(T)$, which turned out to
be semiconductor-like in a wide temperature range above $\sim 350$~K in single crystals \cite{Kar}
and above $\sim 315$~K in polycrystalline materials \cite{300K}.
However, below $\sim 300$~K, $\rho(T)$ transforms into roughly metallic-like dependence \cite{Kar,300K}, and takes a shape resembling that observed for underdoped cuprate HTSCs (cuprates) \cite{Kord,And} and FeAs-based superconductors (Fe-pnictides) \cite{Sad2,SO}.
Eventually, as $T$ decreases, FeSe becomes superconducting (SC) with SC transition temperature $T_{c} \thickapprox 10$~K at ambient pressure \cite{Ross,Sun1,Kar,Sad,Kord}, unexpectedly in a rather narrow range of Se concentration \cite{Pom}.
Interestingly, $T_{c}$ is found to noticeably increase up to 38~K after applying pressure of 9~GPa \cite{H2,Med} or by means of partial substitutions of the Se site with S or Te \cite{Sun1,MiT,4*}.
In some cases, pressure and intercalation being combined result in superconductivity with $T_c = 48$~K \cite{Sun}. Moreover, one-unit-cell-thick FeSe films are reported to demonstrate the critical temperature at about 109~K \cite{EugeneReffive,KordRefsix,KordRefseven} likely pointing out new possibilities for superconductivity in this compound present well above the liquid nitrogen temperature \cite{KordRefnine}.

It has been shown that the maximum in $\rho(T)$, observed for FeSe at about 315-350~K \cite{Kar,300K}, is not related to electron-phonon scattering or spin fluctuations-charge carriers coupling \cite{Kar}, nor charge carriers thermal activation \cite{Bach}.
Most likely, in the range of 315-350~K, the electron band structure of FeSe is reconstructed, which can lead to an increase in the density of charge carriers, $n_f$, and, as a result, to the observed decrease in $\rho(T)$ at lower temperatures (see Ref. \cite{Kar} and references therein). It is important to note that no structural or magnetic transition at about 300-350~K was reported \cite{Kar,300K}.
A poorly recognized structural transition was identified at $T_{s1} = 247$~$(\pm2)$~K \cite{GnRef,Gnesd}, and another well-known structural transition was revealed at $T_{s}\sim 90$~K \cite{Sun1,Kar,Pom,McQ,Bolg,Galluz,Wat1}.
However, unlike Fe-pnictides, for FeSe, the transition at $T_s$ is not accompanied by the corresponding antiferromagnetic (AFM) transition, that is, it seems to be of the nematic type (see Refs. \cite{John,Sad,Kord} and references therein).
There are several pieces of evidence now which suggest charge induced nematicity in FeSe \cite{Sun1,Kar,H2,Kord,Mass,Wat2}.
This type of transition is associated with spontaneous breaking of the symmetry between the $x$ and $y$
directions in the Fe plane, reducing group symmetry of the lattice
from tetragonal to orthorhombic.
This fact reflects the intrinsic electronic instability of FeSe and complex evolution of the
electronic band structure with temperature \cite{Sun1,Kar,H2,John,Sad,Kord,GnRef,Gnesd,McQ,Wat1,Mass,Wat2,Wat}.
Finally, one may conclude that the FeSe compounds are deep in the crossover regime between the BCS and Bose-Einstain-Condensation (BEC) limits \cite{Kas,Mass,Wat2,Wat,Ran}.

The $\rho(T)$ behavior of FeSe differs markedly from that observed for cuprates (such as YBa$_2$Cu$_3$O$_{7-\delta}$ (YBCO), etc.) where $\rho$ is a linear function of temperature in a wide temperature range above the pseudogap temperature $T^*>>T_c$ \cite{Lang,S1}. Below $T^*$, $\rho(T)$ deviates downward from the linearity resulting in appearance of the excess conductivity, $\sigma'$, as a difference between the measured conductivity, $\sigma(T)=1/\rho(T)$, and the extrapolated normal-state linear conductivity, $\sigma_N(T)=1/\rho_N(T)$ \cite{We,Lang,S1}. For YBCO, in a relatively short temperature range $\sim$15~K above $T_c$ \cite{Grb}, $\sigma'(T)$ is well fitted by the classical fluctuation theories of Aslamazov-Larkin (AL) and Hikami-Larkin (HL) (Maki-Thompson (MT) term) (see Refs. \cite{We,S1} and references therein). This is the range of SC fluctuations or the range of fluctuation conductivity (FLC), which is characterized by the nonzero superfluid density, $n_s$, because of formation of fluctuating Cooper pairs (FCPs) above $T_c$ \cite{Cor,Kaw,Yam}. In this temperature range, FCPs behave in a good many way like SC Cooper pairs without long-range order, that is, the pairs which show so-called short-range phase correlations \cite{Ran, EmeryK} and somehow have to obey the BCS theory \cite{DeGen}.

Studies of SC fluctuations have attracted a great deal of attention in the research of cuprate HTSCs (see Refs. \cite{Lang,S1,Grb} and references therein). The main reason for that is related to the nature of the pseudogap (PG), which is known to open below $T^*$, that in underdoped cuprates is much above $T_c$ and may have superconducting nature \cite{S1,TS,Tail,DK}. Unlike cuprates, there is an obvious lack of research on the SC fluctuations in FeSe. As a result, little is known about the existence of the FCPs above $T_c$ and their possible influence on the above mentioned properties of FeSe. Accordingly, the data reported on the possible realization of the PG state in FeSe compounds are rather contradictory \cite{Song,Pall,Sprau,Ross}.

In this paper, we report on the study of SC fluctuations in FeSe polycrystals. We have studied three samples prepared by different techniques \cite{Bolg} to find out the influence of the sample structure and possible defects on FLC. The FCPs contributions to FLC were derived from $\rho(T)$ and $\sigma'(T)$ dependencies and analyzed within a model of local pairs (LPs) \cite{S1}. Our results show that in the range of SC fluctuations, $\sigma'(T)$ near $T_c$ is well described by the three-dimension (3D) AL fluctuation theory. At higher temperatures, a crossover to the state well approximated by the two-dimensional (2D) HL fluctuation theory (2D MT term) occurs. The crossover temperature, $T_0$, allows us to determine the coherence length along the $c$-axis at zero temperature, $\xi_c(0)$, which is several angstroms for all samples. Taking into account the resulting coherence length, a set of additional important samples' parameters was determined in the framework of a simple 2D model of free charge carriers, including $\tau_{\phi} $, which is the lifetime of the FCPs in the region of SC fluctuations \cite{Mats,Sug,S3}. Important implications of these findings are also presented and discussed.

In earlier studies on the excess conductivity in FeSe it was assumed that the normal state occurs at temperatures close to $T_c$ \cite{Kas}. The new approach presented in our work concerns the normal state at temperatures close to room temperature and for this choice we quote a number of arguments. The solution we propose is the only one that leads to results consistent with the results obtained with different methods, authenticating our approach.\\

\indent {\bf 2.\, EXPERIMENT}\\

Three samples of FeSe$_{0.94}$ were prepared by different technique, as described elsewhere \cite{Bolg}. Two samples with nominal compositions FeSe$_{0.94}$+4wt\%Ag (S1) and FeSe$_{0.94}$ (S3) were obtained by partial melting. The third sample, FeSe$_{0.94}$ (S2), was obtained by the solid state reaction (SSR) method \cite{Bolg}. Silver has been widely used as a dopant or additive to improve the microstructure and SC properties \cite{Bolgciteten,Bolgciteeleven,Fab}. The role of Ag addition on structure and SC properties was investigated for polycrystalline Sr$_{0.6}$K$_{0.4}$Fe$_{2}$As$_{2}$ \cite{Bolgcitetwelve} and FeSe$_{0.5}$Te$_{0.5}$ \cite{Bolgcitethirteen}, where a small amount of Ag was found to enter into the crystal structure. Moreover, in our previous works it was established that a small amount of Ag incorporated in the grains of FeSe$_{0.94}$ improves both intra- and intergranular SC properties \cite{Bolg,Galluz}. This results in an increase of $T_c$, the upper critical field, $H_{c2}(0)$, the Ginsburg-Landau parameter, $\kappa$, the critical current density, $J_c$, and pinning energy, as well as in a decrease of the transition width, $\Delta T_c$ \cite{Bolg,Galluz}. It was also found that due to Ag doping the irreversibility field slightly increases and intergranular bonds are improved \cite{Bolgcitefourteen}.

Powder X-ray diffraction measurements were carried out within the range from 5.3$^{\circ}$ to 80$^{\circ}$ 2$\theta$ with a constant step 0.02$^{\circ}$ 2$\theta$ on a Bruker D8 Advance diffractometer with Cu K$\alpha$ radiation and a LynxEye detector. Phase identification was performed with Diffrac$plus$ EVA using ICDD-PDF2 Database \cite{Bolg}. It was found (see Fig.~1 in Ref. \cite{Bolg}) that both undoped samples, S2 and S3, consist mainly of the SC tetragonal phase with traces of the non-superconducting hexagonal phase present in sample S3 prepared by partial melting.
No hexagonal phase was detected in the Ag doped sample, S1, where only 4~wt\% of Ag and trace amounts of Fe were found as impurities. Thus, analogous to the Sn addition in FeSe \cite{Bolgcitefifteen}, Ag seems to help suppress the formation of the hexagonal phase and to increase the amount of the SC tetragonal phase. The lattice parameters of samples prepared by melting (undoped and with addition of Ag) are similar (S3: a = 3.7650~\r{A}, c = 5.5180~\r{A}, and S1: a = 3.7671~\r{A}, c = 5.5193~\r{A}) indicating that Ag is most likely only slightly incorporated in the unit cell \cite{Fab}. The lattice parameters of samples S2, obtained by the SSR method, are: a = 3.7759~\r{A}, c = 5.5180~\r{A}.
It is well known that FeSe has the simplest crystal structure among other Fe-based superconductors, as mentioned above. A unit cell is a tetrahedron with Fe ion in the center and Se in its vertices. The symmetry group is P4/$nmm$ with lattice constants a = 3.768~\r{A} (Fe-Fe distance) and c = 5.517~\r{A} (interlayer distance), with the height of the Se ions over the Fe planes, $z_{Se}$ = 0.2343 ($\simeq 1.45$~\r{A}) \cite{Kar,Sad22,Str}. Thus, the lattice parameters of our samples are consistent with those reported in literature.

Rectangular samples of about $5\times1\times1$ mm$^3$ were cut out of the pressed pellets. A fully computerized setup on the bases of a physical properties measurement system (Quantum Design PPMS-14T) utilizing the four-point probe technique was used to measure the longitudinal resistivity, $\rho_{xx}$, with sufficient accuracy at low frequency ($<20$~Hz) of the measuring current. Silver epoxy contacts were glued to the ends of the sample in order to produce a uniform current distribution in the central region where voltage probes in the form of parallel stripes were placed.
Contact resistances below 1~$\Omega$ were obtained for voltage stripes with a width less than 0.3 mm. We expected to find out the difference in the behavior of the samples, conditioned by the way they are prepared, i.e. the degree of disorder, analyzing the fluctuation conductivity derived from the resistivity results.\\

\begin{figure}[b]
\begin{center}
\includegraphics[width=.60\textwidth]{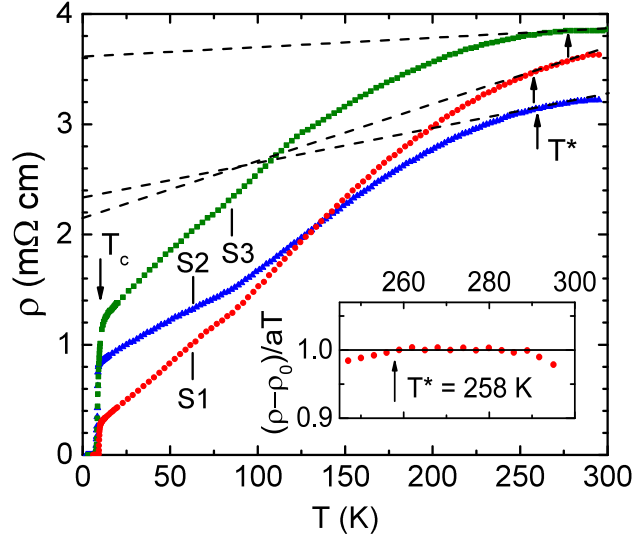}
\caption{Temperature dependencies of resistivity $\rho(T)$ for three samples of
FeSe$_{0.94}$ prepared by different technique (see the text).
Straight dashed lines designate extrapolated $\rho_{N}(T)$. Arrows pointing up
define $T^*$'s for all samples. Insert: ($\rho-\rho_0$)/a$T$ as a
function of temperature for sample S1 (dots), which determines
$T^*$ = 258~$\pm$0.5~K. Straight line is guidance for eye.}
\end{center}
\end{figure}

\indent {\bf 3.\, RESULTS AND DISCUSSION}\\

\indent {\bf 3.1.\, Resistivity}\\

Figure 1 shows the temperature dependencies of the resistivity $\rho(T)=\rho_{xx}(T)$ for all three samples, which exhibit a metallic behavior with positive d$R$/d$T$ until the temperature decreases to $T_c$. By extrapolating the linear part of $\rho(T)$ at the SC transition to $\rho_{ex}=0$, the critical temperature $T_c\equiv T_{c}(\rho_{ex}=0)$ is determined. Since the resistive transitions were quite sharp, especially in the case of S1 \cite{Bolg}, the approach allows to get $T_c$'s values with high accuracy (Table I). As expected, the Ag-doped sample (S1) has the highest $T_c = 9.0~(\pm0.05)$~K and the lowest $\rho(T$=10K)~$ \simeq 270~\mu\Omega$cm. This value is similar to $\rho_{ab}(T$=10K)~$ \simeq 120$ and $500~\mu\Omega$cm, as reported for single crystals in Refs \cite{Yang} and \cite{Lei}, respectively.
Besides that, S1 has the largest residual resistivity ratio RRR = 13.5, defined as R(300 K)/R(10 K). This RRR value is significantly higher than RRR = 9.3, observed for polycrystalline materials obtained by the self-flux method \cite{Song}, which confirms the good quality of the sample structure.
On the other hand, sample S3 has much lover $T_c = 7.8~(\pm0.05)$~K and the highest $\rho(T$=10K)~$\simeq$~1010$~\mu$$\Omega$cm, most likely due to inclusions of the hexagonal phase, as observed by X-ray diffraction \cite{Bolg}, which may perturb the sample structure. The lowest $T_c = 7.7~(\pm0.05)$~K has sample S2 prepared by the SSR method, despite of smaller $\rho(T$=10K)~$\simeq~838~\mu\Omega$cm.
Consequently, relatively small values of RRR, 3.9 for S2 and 3.8 for S3, have been observed, which indicate enhanced disorder in both samples.
Interestingly, despite these differences, samples S1 and S3, obtained by partial melting, have very similar shape of the $\rho(T)$ dependencies. At the same time, the shape of $\rho(T)$ for sample S2, manufactured by the SSR method, changes in a different way and below $T_s\sim$~90~K the slope of the curve becomes smaller than for samples S1 and S3. These observations confirm the expectations that the properties of bulk FeSe substantially depend on the method of manufacturing.
Nevertheless, an obvious kinklike behavior related to the structural transition is observed at $T=T_s$ for all three samples. The structural transition temperature, $T_s$, is usually defined as the temperature at which $d\rho/dT$ takes its minimum value and occurs in the range between 90 and 80~K for FeSe \cite{Sun1,Kar,Bolg}. As mentioned above, in Fe-pnictides the structural transition is followed by the transition to the spin-density-wave AFM state \cite{Sad2}, that may be responsible for the electron pairing and superconductivity \cite{Maz,Wang,Tail}.\\

\indent {\bf 3.2.\, Fluctuation conductivity}\\

As mentioned above, the normal state in cuprates is well defined by the linear $\rho(T)$ dependence at higher temperatures \cite{And,S1,Lang} and is also characterized by the stability of a Fermi surface \cite{Tail,PB,SP,Bad,FerS}.
Nevertheless, below a representative temperature $T^*\gg T_c$, $\rho(T)$ deviates downwards from the linearity and the system goes into the pseudogap state (see Refs. \cite{We,Lang,S1} and references therein) which is characterized by the partial reduction of the electronic density of states (DOS) at the Fermi level \cite{Allo,Kond}, probably due to reconstruction of the Fermi surface \cite{SP,Bad,PB,FerS}.
Usually, $T^*$ is taken at the point where the $\rho(T)$ curve starts to downturn from the high-temperature linear dependence \cite{We,Lang,S1,VS,Oh,Vovk,Kond2}. The deviation results in the emergence of excess conductivity:
\begin{equation}
\sigma '(T) = \sigma(T) - \sigma_N(T)=[1/\rho(T)]-[1/\rho_N(T)],
\label{sigma-t}
\end{equation}
where $\rho_{N}(T)$ is the extrapolated linear normal-state resistivity. Certainly, $\sigma'(T)$ should contain information about fluctuation conductivity due to existence of the SC fluctuations above $T_c$ \cite{We,S1,Lang,VS}.

Unlike cuprates, the normal state in FeSe is still uncertain, despite the seeming simplicity of its structure (see insert in Fig.~3).
Ultimately, the normal state was chosen in the manner shown by the dashed lines in Fig.~1, based on the following considerations.
First, below about 300 K the rearrangement of the electronic band structure is completed, and the FeSe system drops into a new state which is characterised by a metallic type of charge carriers scattering. In this state the Hall coefficient, $R_H$, was found to be almost temperature independent down to about 250~K \cite{Sun1,Kar,H2}. This, in turn, indicates the stability of the Fermi surface, which is the main feature of the normal state of any HTSC, as mentioned above. Additionally, it was found that the field-dependent magnetoresistance, $MR = {[\rho(H) - \rho(0)]/\rho(0)}$), measured under different temperatures, obeys Kohler's law between 250 and 300~K \cite{Naz}, also suggesting the stability of the Fermi surface in this temperature range.
In that case the MR can be successfully scaled by $MR = F(\omega_c\tau) = F\{[\mu_0H/\rho(0)]^2\}$, where $F$ is a function of the cyclotron frequency $\omega_c$ and the scattering time $\tau$, if the scattering rate for charge carriers is equal at all points on the Fermi surface \cite{Zim}. This result looks much more convincing for single crystals \cite{Sun1} and can be considered as a sign of the normal state of the system.
Secondly, like in cuprates, $\rho(T)$ for our samples turned out to be linear, but, as expected, in a relatively short temperature range below $\simeq 290$~K.
For greater certainty, we used a more accurate method of determining $\rho(T)$ linearity exploring criterion $\rho(T) = aT + \rho_0$, where $a$ designates the slope of the extrapolated linear $\rho_{N}(T)$ dependence and $\rho_0$ corresponds to its intercept with $Y$ axis \cite{44}.
Evidently, in the normal state $\rho$ = $\rho_N$ and  [$\rho_N(T) - \rho_0]/aT = 1$. Accordingly, the deviation from unity determines the representative temperature $T^*$ with an accuracy $\pm$1 K \cite{We,S1,44}.

Results of this approach are shown in the insert in Fig.~1, for sample S1 as an example. As one can see, $\rho(T)$ deviates downward from the linearity both above $\simeq 290$~K, due to the beginning of the reconstruction of the electron band structure, and below $T^*\simeq 258$~K, most likely due to a complex interplay between electron-like and hole-like charge carriers, which both are known to be present in FeSe compounds \cite{Sun1,Kar,H2,Sad,Kord,John}. It has been found that both $R_H(T)$ \cite{Sun1,Kar,H2,300K} and thermopower, $S(T)$ \cite{Song}, change their sign several times upon cooling, thereby confirming FeSe to be a compound with two types of charge carriers \cite{Sun1,Kar,H2,Sad,Kord,Song}.
Importantly, we traditionally designate the characteristic temperature as $T^*$, although the reason of changes that begin in $T^*$ is currently not clear \cite{Song,Pall,Sprau,Ross}. Ultimately, $\sigma'(T)$ was calculated using Eq.~(1) and the linear $\rho_N (T)$ dependence shown in Fig.~1. This approach allowed us to get convincing self-consistent results, as will be presented and discussed below.

A significant part of the SC properties of HTSCs, both cuprates \cite{S2} and Fe-pnictides \cite{John,SO}, is determined by the extremely short coherence length in these compounds, $\xi$, which determines the size of the Cooper pairs, both in the $ab$-plane,  $\xi_{ab}$, and along the $c$-axis, $\xi_c$, which at low temperatures is smaller or comparable with the lattice parameter $d$ along this axis (\cite{S1,VS}, and references therein). We consider $\xi(T)=\xi(0)\varepsilon^{-1/2}$, where $\xi(0)$ is the coherence length at $T=0$ and $\varepsilon$ is a reduced temperature, as defined by Eq.~(3) \cite{DeGen}. Near $T_c$, $\xi_c(T)\gg d$, and the FCPs can interact in the entire sample volume, forming a 3D-state of SC fluctuations \cite{We,S1,Lang,Oh,Vovk,Kond2,VS}.
As a result, $\sigma'(T)$ is extrapolated by the standard equation of the Aslamazov-Larkin (AL) theory \cite{AL} with the critical exponent $\lambda=-1/2$, which determines FLC in any 3D superconducting system:
\begin{equation}
\sigma_{AL3D}'=C_{3D}\frac{e^2}{32\,\hbar\,\xi_c(0)}{\varepsilon^{-1\,/\,2}},
\label{3D}
\end{equation}
as shown in Fig.~3 (dashed lines 1), where
\begin{equation}
\varepsilon = (T - T_c^{mf})\,/\,T_c^{mf}
\label{var}
\end{equation}
is a reduced temperature and $C_{3D}$ is a scaling factor used to fit the theory results to the experimental data \cite{S1,Oh,Mats,Sug}. Accordingly, $T_c^{mf}$ is a critical temperature in the mean-field approximation ($T_c^{mf}>T_c$, see Fig.~2) \cite{HL,Xie}, which separates the region of the SC fluctuations ($T \geqslant T_c^{mf}$) from the region of the critical SC fluctuations around $T_c$ (where the SC order parameter $\Delta<kT$ \cite{VarL}), which is not taken into account in the Ginzburg-Landau (GL) theory \cite{DeGen}.

Since above $T_c^{mf}$ $\xi_c(T)$ decreases rapidly with increasing temperature, above the characteristic temperature $T_0 >T_c$ $\xi_c(T)$ will become smaller than $d$ and the Josephson coupling between the FCPs along the $c$-axis will be lost in the entire volume of the sample. But, in the temperature range from $T_0$ to $T_{01}$, $\xi_c(T)$ is still larger than $d_{01}$, where $d_{01}$ is the distance between the inner conducting layers in the unit cell (see Fig.~3(a)), and the inner layers will remain connected by the Josephson coupling forming a 2D fluctuation state  \cite{HL,Xie,VarL,Varl2}.
In this state $\sigma '(\varepsilon)$ is well fitted by the MT term of the HL theory \cite{HL}:
\begin{equation}
\sigma_{MT2D}
'=\frac{e^2}{8\,d\,\hbar}\frac{1}{1-\alpha/\delta}\,ln\left((\delta/\alpha)\,
\frac{1+\alpha+\sqrt{1+2\,\alpha}}{1+\delta+\sqrt{1+2\,\delta}}\right)\,\varepsilon^{-1},
\label{MT}
\end{equation}
as shown in Fig.~3 (solid curves 2) \cite{We,Lang,S2}. In Eq.~(4),
\begin{equation}
\alpha = 2\biggl[\frac{\xi_c(0)}{d}\biggr]^2\,\varepsilon^{-1}
\label{alp}
\end{equation}
is a coupling parameter and
\begin{equation}
\delta = 1.203\frac{l}{\xi_{ab}}
\frac{16}{\pi\,\hbar}\biggl[\frac{\xi_c(0)}{d}\biggr]^2\,k_B\,T\,\tau_{\phi}
\label{TM}
\end{equation}
is the pair-breaking parameter. Accordingly, $l$ is the mean-free path and $\tau_{\phi}$ is the phase relaxation time of the FCPs, which is determined as follows:
\begin{equation}
\tau_{\phi}\beta\,T={\pi\hbar}/{8k_B\varepsilon}=A/\varepsilon,
\label{tau}
\end{equation}
with $A=2.998\cdot 10^{-12}$ sK \cite{S2}. Equation (7) was obtained assuming $\alpha = \delta$, which according to HL theory occurs at the temperature where the AL(3D)-MT(2D) crossover appears. Factor $\beta = 1.203(l/\xi_{ab}(0))$ takes into account the approximation of the clean limit $(l>\xi$) \cite{S1,S2}. It is worth to note that this approximation works only in well structured samples \cite{S1,VS}.
In HTSCs containing defects the contribution of MT fluctuations vanishes and above $T_0$ FLC is described by the Lowrence-Doniach model \cite{Oh,HL}.

As mentioned above, one more characteristic temperature of the FLC analysis is $T_{01}>>T_c$, at which the $\sigma '(\varepsilon)$ experimental results deviate downward from the theory of MT fluctuations (Fig.~3, $ln\varepsilon_{01}$). This is because above $T_{01}$, which actually limits the range of SC fluctuations from above,
$\xi_c(T)$ is lower than the distance $d_{01}$ \cite{We,Lang,S2,Xie}. This, in turn, means that at $T>T_{01}>T_0$ all charge carriers, including the FCPs, are believed to be mainly confined within conducting As-Fe-As layers (Fe-pnictides) \cite{SO} or CuO$_2$ planes (cuprates) \cite{S2}, forming the quasi-2D conducting system \cite{S1,44,Varl2,Xie,Haus,L,Eng}. At those temperatures, because $\xi_c(T) < d_{01}$, there is no direct interaction even between the internal conducting planes and now $\sigma'(\varepsilon)$ does not obey any conventional fluctuation theory. In the case of cuprates, this fact leads to a very unusual behavior of HTSCs observed over the wide temperature range from $T_c$ to $T^*$, which is just called the pseudogap temperature \cite{Cor,Kaw,Yam,Tal,Kord2}.
Thus, HTSCs appear to be quasi two-dimensional systems in the wide temperature range well above $T_c$.
In accordance with the theory, Gaussian fluctuations of the order parameter prevent any phase coherency in 2D compounds \cite{DK,Haus,L,Eng}.
Consequently, the SC critical temperature of an ideal 2D metal is found to be zero (Mermin-Wagner-Hoenberg theorem) and a finite value is obtained only when three-dimensional effects are taken into account (see Refs. \cite{S1,VS,Haus,L,Eng,Tal,Kord2} and references therein). That is why in HTSCs near $T_c$, where $\xi_c(T)>d$, the three-dimensional state is realized and FLC is extrapolated by the standard 3D equation of the AL theory (Eq.~(2)). This scenario is recognized as valid in the case of both cuprates \cite{S1,S2} and Fe-pnictides \cite{We,SO,S1,Salem,Rey,Ahmad,Asiyaban}.

In order to study the SC fluctuations in FeSe and examine their role in the formation of the SC state, it is proposed to calculate  $\sigma'(T)$ with Eq.~(1) and plot it as a function of $\varepsilon$, and then fit the experimental data to the results predicted by fluctuation theories of AL and HL (2D MT term). But the first step in using this approach is to search for the mean-field critical temperature $T_c^{mf}$, which determines the reduced temperature $\varepsilon$ (Eq.~(3)).
From Eq.~(2) it is clear that near $T_c^{mf}$, $\sigma '(T)$ diverges as $\varepsilon^{-1/2}$. Consequently, $\sigma '^{-2}(T)$ becomes zero when $T = T_c^{mf}$, as $\varepsilon = (T - T_c^{mf})\,/\,T_c^{mf}$ (Eq.~(3)) and Refs. \cite{Oh,Bis}). Hence, in order to find $T_c^{mf}$, $\sigma '^{-2}$ is plotted as a function of temperature for all samples studied, as shown in Fig.~2. The intersection of the linear part of each $\sigma '^{-2}(T)$ curve with $T$ axis determines the values of $T_c^{mf}$, which are listed in Table~I. In Fig.~2, the critical temperature (taken at $\rho_{ex}=0$, as described in Ch.~3.1), $T_c$, the Ginzburg temperature (taken at the deviation from the linearity), $T_G$, and the 3D-2D crossover temperature, $T_0$, are also shown. $T_G$ is a temperature which is generally accounted for by the Ginzburg criterion, which is related to the breakdown of the mean-field GL theory to describe the SC transition \cite{GL}. This criterion is identified down to the lowest temperature limit for the validity of the Gaussian fluctuation region \cite{DeGen,GL,Kap}. In Fig.~3, $T_G$ corresponds to the value marked ln($\varepsilon_G$).

\begin{figure}[t]
\begin{center}
\includegraphics[width=.58\textwidth]{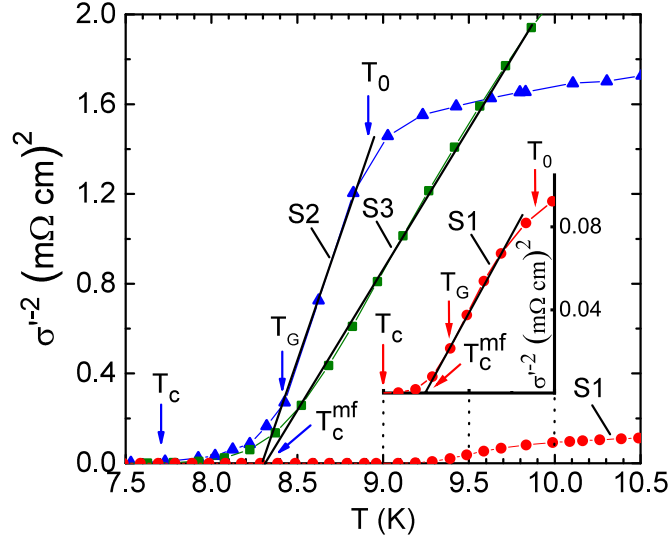}
\caption{Temperature dependencies of $\sigma^{\prime-2}$, which allow to determine the mean-field critical temperature, $T_c^{mf}$,
for FeSe$_{0.94}$ samples: S1 - dots, S2 - triangles and S3 - squares. The critical temperature, $T_c$, the Ginzburg temperature, $T_G$, and the 3D-2D crossover temperature, $T_0$, are also shown, for clarity for samples S1 and S2 only. Insert: $\sigma^{\prime-2}(T)$ dependence for sample S1, plotted on an enlarged scale.}
\end{center}
\end{figure}

It is worth noting that the linear part of every curve in Fig.~2, extrapolated by a straight line at $T<T_0$, corresponds to the temperature range where 3D AL fluctuations are present \cite{We,S2,HL,Xie,VS}. Above $T_0$, where $\xi_c(T)<d$, the data deviate from the line, suggesting the contribution of 2D MT fluctuations \cite{S1,HL,Xie}, as will be discussed below in detail. In Fig.~3 this temperature corresponds to the value marked ln($\varepsilon_0$). Evidently, at $T_0$ the coherence length $\xi_c(T_0)=\xi_c(0)\varepsilon_{0}^{-1/2}$ is expected to amount to $d$ \cite{S2,Mats,Sug,HL}, which yields:
\begin{equation}
\xi_c(0) = d\,\sqrt\varepsilon_0,
\end{equation}
and allows one to determine $\xi_c(0)$, which is one of the important parameters in the FLC analysis \cite{S1,Lang,HL,Xie,VS}.
Importantly, it was found that the temperature of the 3D-2D crossover at $ln(\epsilon_0)$ (Fig.~3), and hence the value of $\xi_c(0)$, is independent on the choice of the normal state from which the linear part of $\rho(T)$ is extrapolated to lower temperatures, so both $T_0$ and $\xi_c(0)$ represent reliable values.

The excess conductivities of FeSe$_{0.94}$ samples, derived from the resistivity measurements by means of Eq.~(1), are plotted in Fig.~3 as a function of $\varepsilon$ in double logarithmic scale. The experimental results (symbols) were fitted by the theoretical curves according with Eq.~(2) (dashed lines) and Eq.~(4) (solid curves). Results obtained for the Ag-doped sample (S1, dots in Fig.~3(a)) will be considered in details as an example. In agreement with the performed considerations, above $T_G=9.4$~K (from ln$\varepsilon_G = -4.19$) and up to $T_0=9.9$~K (from ln$\varepsilon_0 = -2.73$), ln$\sigma '$ versus ln$\varepsilon$ is expectedly well fitted by the 3D fluctuation term of the AL theory (Eq.~(2)) with $\xi_c(0) = 2.8~(\pm0.05)$~{\AA}, as determined by Eq.~(8) and with $C_{3D} = 3.04$ (Table~I). By analogy with cuprates, to find $\xi_c(0)$ we used $d=11.0$~{\AA} and put $\varepsilon_0 = 0.065$, as obtained from ln$(\varepsilon_0) = -2.73$. It should be emphasized that, despite the fact that FeSe lattice parameters are considered to be well established \cite{Kar,Sad2,Bolg,Sad22}, the value of the lattice parameter of the $c$-axis used in the FLC analysis is still somewhat doubtful.

There is definite contradiction between stated in Chapter~2 $c=5.52$~{\AA} and taken by us $d = 2c = 11.0$~{\AA}. As well as in all other Fe-based superconductors, the charge carrier transport in FeSe is assumed to be realized within the conducting Se-Fe-Se layers \cite{Kord,Sad,Sad2}. However, any such a layer alone can't be superconducting and one needs more layers connected by SC correlations along the $c$-axis to observe superconductivity: two layers to obtain the 2D state and at least three layers to observe the 3D state near $T_c$, which is necessary to ensure a SC transition \cite{Xie,Khas}. Thus, it is clear that $d=11.0$~{\AA} determines just the $c$-axis lattice parameter of the "superconducting unit cell", whereas $c=5.52$~{\AA} corresponds to the distance between the adjacent Fe layers. In Ref. \cite{Kord} it is claimed that for FeSe-based superconductors, the photon energy dependence of the experimentally measured Fermi surface supports the idea of doubling of the unit cell along the $c$-axis, thus confirming our approach to take $d=2c=11.0$~{\AA}.

For high-$T_c$ superconductors, the coherence length in the $ab$-plane is $\xi_{ab}(0)\approx$ (10-15)$\xi_{c}(0)$ \cite{VS,S2,Oh,Mats,Sug}.
Taking $\xi_c(0)=2.8$~{\AA}, as derived above (Table~I, S1), and $\xi_{ab}(0)=34$~{\AA}, as obtained from magnetic measurements in our previous work \cite{Bolg}, we have $\xi_{ab}(0)/\xi_{c}(0) \simeq 12$, which implies our estimation of $\xi_{c}(0)$ to be correct.
Moreover, $\xi_c(0)=2.8$~{\AA}, found for our FeSe sample, is approximately the same as obtained for EuFeAsO$_{0.85}$F$_{0.15}$ with $T_c = 11$~K ($\xi_{c}(0)=2.84$~{\AA}) \cite{SO} and about 2 times larger than that received for SmFeAsO$_{0.85}$ with $T_c = 55$~K ($\xi_{c}(0)=1.40$~{\AA}) \cite{Sm}. This is not surprising, since we assume that $\xi_{c}(0)\sim \hbar v_F/\pi\Delta(0)\sim \hbar v_F/\pi k_B T_c$, with regard to the BCS ratio $2\Delta(0) \sim k_BT_c$ \cite{DeGen}. In other words, the lower $T_c$ the higher both $\xi_c(0)$ and, respectively, $\xi_{ab}(0)$, in accordance with our results. Both $\xi_c(0)$ and the corresponding $\xi_{ab}(0)$, as well as other parameters, have been found in the same way for other samples and are summarized in Table~I.

Figure~3 shows that above $T_0$ the measured $\sigma '(\varepsilon)$ deviates noticeably upward from the linear dependence of the 3D AL theory (Eq.~(2), dashed lines 1), which indicates the appearance of 2D MT fluctuations. So the 3D state is lost, however $\xi_c(T) \gtrsim d_{01}$ and the inner conducting layers are still connected by the Josephson coupling forming a state with 2D fluctuations as explained above and in Refs. \cite{HL,Xie,VarL}. As a result, $\sigma '(\varepsilon)$ is well approximated by Eq.~(4) (solid curves 2 in Fig.~3) up to $T_{01}$, which for sample S1 is equal to 19~K (from ln$(\varepsilon_{01}) = 0.05)$.
It is worth to emphasize that $T_{01}=19$~K is approximately double $T_c$, in good agreement with results reported in Refs. \cite{Kas,Naid}.

\begin{figure}[b]
\begin{center}
\includegraphics[width=.55\textwidth]{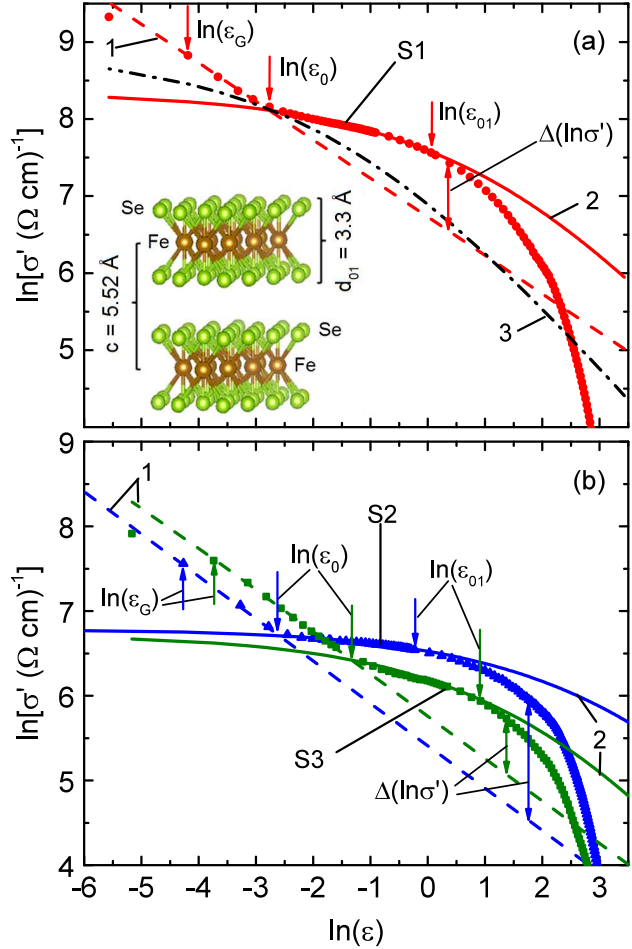}
\caption{Excess conductivity, $\sigma'$, versus reduced temperature, $\varepsilon$, for (a) S1 and (b) S2, and S3 samples of FeSe$_{0.94}$, compared with the results obtained in the frame of the superconducting fluctuation theories: 3D AL (1, dashed lines), 2D MT (2, solid curves) and, for sample S1, 2D MT in the nonmagnetic limit (3, dash-dotted curve). Vertical arrows designate logarithms of reduced temperatures (see Eq.~(3)) corresponding to $T_G$, $T_0$ and $T_{01}$. Double vertical arrows show the enhancement of the fluctuation conductivity due to 2D MT contribution, $\Delta$ln$(\sigma')$. Insert to (a): schematic of the crystal structure of FeSe.}
\end{center}
\end{figure}

\begin{table}[tbp]
\caption {Parameters of FeSe$_{0.94}$ samples (S1-S3) obtained from the fluctuation conductivity analysis:
$\rho$(25K)~- sample resistivity at 25~K;
$T_c$~- critical temperature;
$T_c^{mf}$~- mean-field critical temperature;
$T_G$~- Ginzburg temperature;
$T_0$~- 3D-2D crossover temperature;
$T_{01}$~- temperature at which the data deviates downward from the theory of MT fluctuations;
$\Delta T_{fl}$~- range of the superconducting fluctuations, $\Delta T_{fl} = T_{01} - T_G$;
$C_{3D}$~- scaling factor;
$\xi_c(0)$~- coherence length along the $c$-axis at $T=0$;
$\xi_{ab}(0)$~- coherence length along the $ab$-plane at $T=0$;
$d_{01}$~- distance between the inner Se sheets in the FeSe conducting layer;
$\Delta$ln$\sigma'$~- enhancement of the excess conductivity due to 2D MT fluctuations.}\

\centering
\begin{tabular}{|c|c|c|c|c|c|c|c|c|c|c|c|c|}
\hline
Sample & $\rho$(25K) & $T_c$ & $T_c^{mf}$ & $T_G$ & $T_0$ & $T_{01}$ & $\Delta T_{fl}$ & $C_{3D}$ & $\xi_c(0)$ & $\xi_{ab}(0)$ & $d_{01}$ & $\Delta$ln$\sigma'$ \\
& (m$\Omega$cm)& (K) & (K) & (K) & (K) & (K) & (K) & (-) & ({\AA}) & ({\AA}) & ({\AA}) & (-) \\
[0.5ex] \hline S1 & 0.50 & 9.0 & 9.25 & 9.4 & 9.9 & 19 & 9.6 & 3.04 & 2.8 & 34 & 2.7 & 0.90 \\
[0.5ex] \hline S2 & 1.00 & 7.7 & 8.31 & 8.4 & 8.9 & 15 & 6.6 & 0.87 & 2.9 & 43 & 3.2 & 1.42 \\
[0.5ex] \hline S3 & 1.47 & 7.8 & 8.32 & 8.5 & 10.5 & 29 & 20.5 & 2.40 & 5.6 & 36 & 3.6 & 0.66 \\
\hline
\end{tabular}
\label{tab:sample-values1}
\end{table}

Taking into account the considerations presented, one may conclude that $\xi_c(T_{01}) = d_{01}$ and, in accordance with $\xi_c(T_{01})=\xi_c(0)\varepsilon_{01}^{-1/2}$ and Eq.~(8), the following equality is met: $\xi_c(0) = d\sqrt{\varepsilon_0} = \xi_c(T_{01})\sqrt{\varepsilon_{01}} = d_{01}\sqrt{\varepsilon_{01}}$. As far as both $\varepsilon_0$ and $\varepsilon_{01}$ have been determined, and besides $d=11.0$~{\AA}, simple algebra yields: $d_{01} = d\sqrt{\varepsilon_0/\varepsilon_{01}} \simeq 2.7$~{\AA} for S1 (Table~I).
In Fe pnictides, $d_{01} \approx 3$~{\AA} is the distance between As atoms in the conducting As-Fe-As layers \cite{Sm,Ni,J}. By analogy with the Fe pnictides, in FeSe $d_{01}$ is expected to be the distance between Se atoms in the conducting Se-Fe-Se layers. Thus, $d_{01} = 2.7$~{\AA} is consistent with the crystal structure shown in the insert of Fig.~3. It is important to notice that in order to get the proper MT fit we have to put $d_{01}$ instead of $d$ into Eqs.~(4-6) and $\varepsilon_{01}$ instead of $\varepsilon$ into Eq.~(7) to calculate $\tau_{\phi}\beta T$. Otherwise, using Eq.~(4) we obtain the curve which does not fit the experimental results (dash-dotted curve in Fig.~3(a)) and characterizes non-magnetic cuprates \cite{S1,S2}.
Therefore, this means that above $T_0$ FeSe exhibits an increased contribution of 2D FLC, which seems to be characteristic for magnetic superconductors \cite{We,SO,Sm}, where fluctuations of magnetic moments are present \cite{Ross}. However, as in cuprates, $\delta \approx 2$ (Eq.~(6)), which implies that all FLC parameters are defined properly \cite{S1,S2}. Finally, using Eq.~(7) and assuming that $\tau_{\phi} \sim 1/T $ \cite{Mats} we get $\tau_{\phi}\beta T \simeq 28.5\cdot10^{-13}$~sK, and for $T=25$~K, we obtain $\tau_{\phi}\beta \simeq 1.14\cdot10^{-13}$~s (S1, Table~II).

The corresponding ln$\sigma'$(ln$\varepsilon$) dependencies obtained for samples S2 and S3 are also shown in Fig.~3. Near $T_c$, both dependencies are well fitted by Eq.~(2) of the 3D AL approach, which implies the formation of the three-dimensional state at these temperatures. Above $T_0$, which in the Figure is denoted as ln$\varepsilon_0$, the superconductors drop into the 2D state, and ln$\sigma'$(ln$\varepsilon$) is now well fitted by the 2D MT term according to Eq.~(4).
Finally, above $T_{01}$ (ln$\varepsilon_{01}$ in the Figure), the ln$\sigma'$(ln$\varepsilon$) dependencies deviate downward from the prediction of the theory demonstrating behavior similar to that observed for sample S1. However, the ln$\sigma'$(ln$\varepsilon$) curves show several differences, which are believed to arise from different samples' compositions and/or preparation methods, as it was also concluded in Ref. \cite{Ross}.

\begin{figure}[b]
\begin{center}
\includegraphics[width=.59\textwidth]{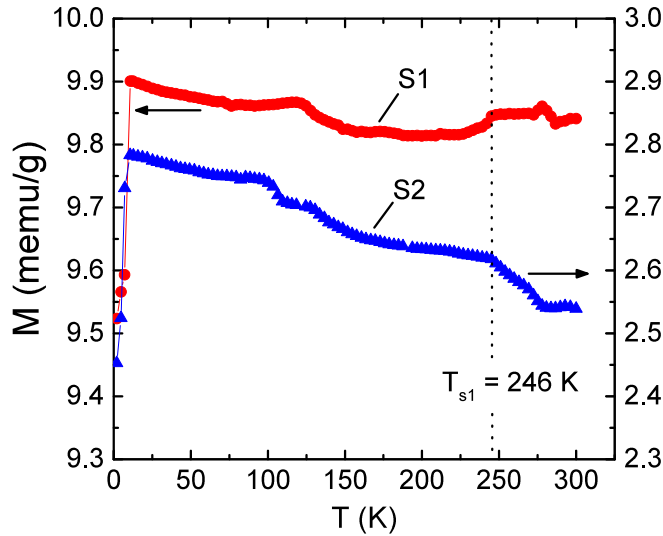}
\caption{Magnetization as a function of temperature for samples S1 and S2, measured at $H=20$~Oe. The dashed line indicates the temperature $T_{s1}$, where an anomaly in the $M(T)$ dependence occurs.}
\end{center}
\end{figure}

Analyzing the transport properties of our samples we found that the resistivity of sample S3 is at low temperatures about three times larger than that of sample S1 (see Fig.~1 and Table~I), however both samples were prepared similarly by partial melting. At the same time, the fluctuation conductivity of S3 is about three times smaller than that of S1 (see Fig.~3). Despite the fact that S3 contains traces of the non-superconducting phases, it shows a very good fit of the experimental results to relations predicted by the theory in both 3D and 2D fluctuation regions, as presented in Fig.~3. Importantly, all three characteristic temperatures, $T_G$, $T_0$ and $T_{01}$, are clearly marked in the ln$\sigma'$(ln$\varepsilon$) curve.
Comparing to sample S1, $T_{0}$ and $T_{01}$ for S3 are significantly shifted towards higher temperatures, probably due to the increased range of 3D fluctuations, which is $T_0-T_G = 0.5$ and 2.0~K, for S1 and S3, respectively. This leads to extremely high $T_{01} \simeq 29$ K (from ln$\varepsilon_{01} = 0.92$ in Fig.~3), showing that in S3 the range of SC fluctuations extends to temperatures about 3.7 times larger than $T_c = 7.8$~K. Thus, one may conclude that various inhomogeneities, for example, traces of non-superconducting phases, not only increase the resistivity but also the range of SC fluctuations above $T_c$. It seems to be reasonable if one assumes the existence of gossamer-like high-temperature superconductivity, which could appear in a form of disconnected superconducting islands, as was observed in FeSe by Sinchenko, $et~al.$, \cite{Goss}. With this approach, it is believed that inhomogeneous superconductivity in FeSe may appear below $\sim 35$~K, much above $T_c = 8$~K, most likely in the form of SC fluctuations, as we have found for our samples.

In the case of sample S2, prepared by the SSR method, the excess conductivity in the region of 3D AL fluctuations is smaller than that determined for S3 (see Fig.~3), however it is markedly larger in the region of 2D MT fluctuations resulting in the largest $\Delta ln\sigma'$, which is characteristic of magnetic superconductors \cite{We,SO,Sm}. Thus, for sample S2, one may conclude, that the increased contribution of 2D fluctuations is due to enhanced magnetic interaction.
For greater confidence, we have measured the temperature dependence of the magnetisation, $M(T)$, for samples S1 and S2, and the results are shown in Fig.~4. For both $M(T)$ curves, a clear anomaly has been observed at $T_{s1}\simeq 246$~K, which corresponds to the mentioned less known structural or (less possible) magnetic transition present in FeSe compounds \cite{GnRef,Gnesd}. Interestingly, close to this temperature, the Hall resistance changes its sign to negative \cite{Sun1}. All these facts provide an additional argument to choose the particular state of FeSe above $T_{s1}$ as the normal state of the sample, which is the reference state for calculation of the excess conductivity.

There are several differences in the $M(T)$ dependencies observed for samples S1 and S2. Below $T_{s1}$, the magnetization for S1 changes more subtle than for S2. Namely, the increase in $M(T)$ in the range from $T_{s1}$ down to $T_c$ is for S2 about 3 times larger than for S1. This observation is consistent with the enhanced magnetic fluctuations in S2, as revealed in the FLC analysis, and supports the statement that the samples properties strongly depend on the sample preparation method. The enhanced magnetic fluctuations may lead to a rather specific shape of the experimental ln$\sigma'$(ln$\varepsilon)$ curve, which for sample S2 is almost parallel to the $x$ axis in the 2D fluctuation range (Fig.~3).

Fitting the experimental ln$\sigma'$(ln$\varepsilon)$ data for sample S2 by Eq.~(4) allows us to determine ln$\varepsilon_{01} = -0.20$ (Fig.~3, $T_{01} \simeq 15$ K), and this yields $d_{01}= \xi_c(0)/\sqrt{\varepsilon_{01}} \simeq 3.2$~{\AA} in good agreement with the width of the Se-Fe-Se layer (see insert in Fig.~3). Here, $\xi_c(0) = 2.9~(\pm 0.05)$~{\AA} has been independently determined by the 3D-2D (AL-MT) crossover temperature $T_0 \simeq 8.9$~K (from ln$\varepsilon_0 = -2.6$), in a similar way as for samples S1 and S3. This value of $\xi_c(0)$ results in a reasonable anisotropy of the coherence length, expressed by the ratio $\xi_{ab}(0)/\xi_c(0) = 43/2.9 \simeq 15$.
Smaller values of the anisotropy were found from the measurements of $H_{c2}$, namely $\sim 10$ \cite{Farrar}, $\sim 7$ \cite{Shiogai} and $\sim 2$ \cite{Her}, indicating that the exact values of the anisotropy and their interrelationships are still under discussion.
The range of the SC fluctuations extends to the temperature $T_{01} \simeq 15$~K, which gives a ration $T_{01}/T_c \simeq 1.9$ and is still about two times higher than $T_c$, comparable to the results obtained for S1 ($T_{01}/T_c \simeq 2.1$) and S3 (3.7). Finally, taking into account the characteristic temperatures we discussed, the parameters $\xi_c(0)$, $d_{01}$, $\Delta$ln$\sigma'$ and $\tau_{\phi}\beta T$ were determined for all samples and are listed in Tables~I and II.\\

\begin{table}[tbp]
\caption {Parameters of FeSe$_{0.94}$ samples (S1-S3) and the well-structured YBCO film (F1) obtained from the phase relaxation time analysis:
$\rho$(25K)${\cdot}C_{3D}$~- sample resistivity at 25~K multiplied by scaling factor;
$n_\text{f}$~- charge-carrier density;
$\mu_H$~- mobility of Hall carriers;
$l(25$K)~- mean free path of charge carriers at 25~K;
$v_{F}$~- Fermi velocity;
$m^{*}/m_0$~- relative effective charge-carrier mass;
$\tau(25$K)~- scattering time (transport) at 25~K for normal charge carriers;
$\tau_{\phi}{\beta}$~- parameter estimated from the fluctuation conductivity analysis;
$\beta$~- factor which takes into account the approximation of the clean limit;
$\tau_{\phi}(25$K)~- phase relaxation time at 25~K for fluctuating Cooper pairs.
For sample F1, all parameters are given at 100 K.}\

\centering
\begin{tabular}{|c|c|c|c|c|c|c|c|c|c|c|}
\hline
Sample & $\rho$(25K)${\cdot}C_{3D}$ & $n_f$ & $\mu_H$ & $l(25$K) & $v_F$ & $m^{*}/m_0$ & $\tau(25$K) & $\tau_{\phi}{\beta}(25$K) & $\beta(25$K) & $\tau_{\phi}(25$K) \\
& ($\mu\Omega$cm)& (10$^\text{21}$cm$^\text{-3}$) & (cm$^\text{2}$/Vs) & ($10^{-8}$cm) & $(10^7$cm/s) & (-)  & $(10^{-13}$s) & $(10^{-13}$s) & (-) & $(10^{-13}$s) \\
[0.5ex] \hline S1 & 1520 & 0.57 & 7.2 & 9.4 & 0.19 & 12.2 & 0.50 & 1.14 & 0.33 & $3.45\pm 0.1$ \\
[0.5ex] \hline S2 & 870 & 0.57 & 12.6 & 16.4 & 0.32 & 7.3 & 0.52& 1.47 & 0.46 & $3.20\pm 0.1$ \\
[0.5ex] \hline S3 & 3530 & 0.57 & 3.1 & 4.1 & 0.17 & 13.3 & 0.24& 0.48 & 0.14 & $3.42\pm 0.1$ \\
[0.5ex] \hline F1 & 147(100K) & 2.55 & 16.6 & 48.5 & 1.17 & 4.7 & 0.42 & 15.1 & 4.5 & $3.35\pm 0.1$ \\
\hline
\end{tabular}
\label{tab:sample-values2}
\end{table}

\indent {\bf 3.3.\, Phase relaxation time comparative analysis}\\

Having determined the FLC parameters it would be interesting to consider physical implication of the short coherence length $\xi_{ab}(0)$ within a simple two-dimensional free-carrier picture \cite{Mats,Sug,S3}. The approach allows to derive a set of additional important samples' parameters including $\tau_{\phi}$, which is actually the lifetime of the FCPs in the range of SC fluctuations. In HTSCs all parameters, including $\tau_{\phi}$ and $R_H$, are functions of temperature. In the literature, the corresponding samples' parameters are usually calculated at temperatures above but not far from $T_c$, e.g. in YBCO at $T=$~100~K \cite{S3}. By analogy, we will perform our calculations at $T=$~25~K, which is just above the SC fluctuation region in FeSe$_{0.94}$. Since $\xi_{ab}(0)$ has been already obtained \cite{Bolg} (Table~I), to calculate $\beta=1.203l/\xi_{ab}(0)$, and then to get $\tau_{\phi}$ using the value of $\tau_{\phi}\beta T$ (Table~II), we need to know the mean free path, $l$, which can be found from $R_H(T)$ \cite{Mats}. For FeSe single crystals, corresponding $R_H(T)$ is reported in Refs. \cite{Sun1,H2}. However, most likely due to different ways of the single crystals preparation, $R_H$(25K)~$\approx -7\cdot10^{-9}$ and $-15\cdot10^{-9}$~m$^3$/C have been obtained in \cite{Sun1} and \cite{H2}, respectively. Eventually, we make use of average $R_H$(25K)~$=-11\cdot10^{-9}$ m$^3$/C to roughly estimate all required parameters. As an example, we will analyze in details results obtained for sample S1, which has the best structural and superconducting properties \cite{Bolg,Galluz}.

In systems with a complicated charge-carrier scattering mechanism the charge-carrier density may be expressed as $n_f = r(1/eR_H)$ \cite{Sze}. Here $e$ is the electron charge and the coefficient $r=<\tau^2>$/$<\tau>^2$, where $\tau$ determines the scattering mechanism in the normal state, as described before. Then, using the aforementioned value of $R_H$(25K)~$=-11\cdot10^{-9}$ m$^3$/C, we obtain for sample S1: $n_f=0.57\cdot10^{21}$~cm$^{-3}$ ($r=1$). This value of $n_f$, together with the corrected resistivity $\rho=\rho$(25K)${\cdot}C_{3D}=500\cdot3.04=1520~\mu\Omega$cm \cite{Mats,Sug}, results in the mobility of Hall carriers $\mu_H=r/(\rho n_f e) \simeq 7.2$ ~cm$^2$/V$\cdot$s. Hence, the mean free path ($l=v_F\cdot\tau$) at 25~K is estimated, $l(25$K$)\simeq 9.4$~{\AA}, from the relation $l=(\hbar\mu_H/e)(2\pi n^*_f)^{1/2}$ with $n^*_f = n_fd = 0.57\cdot10^{21}$~cm$^{-3}\cdot11.0\cdot10^{-8}$~cm~$\simeq 0.63\cdot10^{14}$~cm$^{-2}$, similarly as it was done in Ref.~\cite{S3}.
In the general theory of superconductivity, $\xi_0 \simeq \hbar v_F/(\pi\Delta(0))$ \cite{DeGen}, where $\Delta(0)$ is the SC order parameter at $T=0$~K. Since for HTSCs, $\Delta(0) \sim \Delta^*(T_c)$ \cite{S1,Yam,Kord2,Sta}, where $\Delta^*$ is a pseudogap parameter, taking $2\Delta^*(T_c)/k_BT_c = 3.0$ (S1) (4.6 for S2, and 3.0 for S3) \cite{S-New} and setting $\xi_0=\xi_{ab}(0)$ \cite{Mats} we have obtained the Fermi velocity $v_F \simeq 0.19\cdot10^7$~cm/s, the scattering time (transport) for normal carriers $\tau$(25K)~$=l/v_F \simeq 0.50\cdot10^{-13}$~s,
and an effective carrier mass of $m^*/m_0=(\rho l)(ne^2)/v_F m_0 \simeq 12.2$.
Finally, we have derived $\beta(25$K$)=1.203{\cdot}l/\xi_{ab} \simeq 0.33$ and using $\tau_{\phi}$(25K)$\beta = 1.14\cdot10^{-13}$~s (found in the FLC analysis) we have obtained $\tau_{\phi}$(25K)$~\simeq 3.45\cdot10^{-13}$~s, surprisingly in good agreement with the result for the well-structured YBCO film with $T_c = 87.4$~K, in Table~II marked as F1 \cite{S3}. Using the same approach as above, the set of corresponding parameters was calculated for other two samples, S2 and S3, and all these parameters are listed in Table~II.

The result, that the transport mean free path is noticeably smaller than the in-plane coherence length (e.g., for S1, $l/\xi_{ab} \simeq 0.28$) implies that, unlike cuprates, FeSe is the type-II superconductor in the dirty rather than in the clean limit. Importantly, most of the parameters derived for the FeSe samples differ markedly from those observed for the well-structured YBCO film (Table~II, sample F1). Most of them are smaller than those found for the film, namely: $n_f$(F1)/$n_f$(S1)~$ \simeq 4.5$, $\mu_H$(F1)/$\mu_H$(S1)~$ \simeq 2.3$, $l$(F1)/$l$(S1)~$ \simeq 5.2$, and $v_F$(F1)/$v_F$(S1)~$ \simeq 6.2$.
Note, that Fermi velocity $ v_F \approx 0.9 \cdot 10^7$ cm/s, which is large but still lower than that observed for the YBCO film (F1), has been reported for an FeSe single crystal with $T_c = $ 9.0 K \cite{Sun1}. At the same time, the effective mass of the charge carriers for sample S1, $m^*$(S1)$/m_0 \simeq 12.2$, which, importantly, does not depend on $C_{3D}$, turned out to be markedly larger than that for the YBCO film, namely, $m^*$(S1)$/m_0/m^*$(F1)$/m_0 \simeq 2.6$.
These results indicate a profound difference in transport properties between FeSe and cuprates. But, despite all the differences discussed above, a rather interesting and surprising result has been obtained for the transport relaxation time of normal carriers ($\tau$) and, more importantly, the phase relaxation time of FCPs ($\tau_{\phi}$). Both $\tau$ and $\tau_{\phi}$ turned out to be very similar for the FeSe samples and the YBCO film (Table~II).

Results concerning the relaxation time allow us to infer that the SC fluctuations may have similar nature and roughly a similar temperature range ($\Delta T_{fl} = T_{01} - T_G$) in different types of HTSCs. For our FeSe samples (Table~I), $\Delta T_{fl}$ extends from 9.6~K (S1) to 20.5~K (S3), which corresponds well with $\Delta T_{fl} \simeq 9.2$~K in the YBCO film (F1) \cite{S1,S2} and with $\Delta T_{fl}$ changes from 7 to 16~K in PrBCO-YBCO superlattices and heterostructures \cite{We}.
Accordingly, $\Delta T_{fl} = 7$~K and $\Delta T_{fl} = 23$~K are reported in Ref. \cite{Grb}, for the nearly optimally doped ($T_c = 89$~K) and deeply underdoped ($T_c = 57$~K) YBCO single crystals, respectively. Thus, we conclude that the SC fluctuations ranges from about 10 to about 20~K above $T_c$ in many different HTSCs. It has to be noted that sample S2, prepared by the SSR method, somehow drops out of the common picture. Indeed, it has the largest $\mu_H$, $l$, and $v_F$, but the lowest resistivity and effective carrier mass of $m^*/m_0 \simeq 7.3~$. However, despite all the differences and the fact that S2 has the smallest $\Delta T_{fl} \simeq 6.6$~K, the phase relaxation time $\tau_{\phi}$(25K)~$\simeq 3.2 \cdot 10^{-13}$~s has been determined, in good agreement with the results obtained for other samples.\\

\indent {\bf  CONCLUSION}\\

For the first time, the fluctuation conductivity, $\sigma'$, of FeSe$_{0.94}$ samples prepared by different techniques has been analyzed within the local pair model showing that in the temperature range from $T_c$ to $T_0$, $\sigma'(T)$ is described by the 3D fluctuation theory of Aslamazov-Larkin (AL) \cite{AL} and, in the range from $T_0$ to $T_{01} >> T_c$, by the 2D fluctuation theory of Hikami-Larkin (MT term) developed for HTSCs \cite{HL}.
Thus, as well as in cuprates, $T_{01}$ designates the range of SC fluctuations above $T_c$, where the fluctuating Cooper pairs behave in good many ways like conventional superconducting pairs but without long-range ordering \cite{Ran,PB,Tail,EmeryK}.
Importantly, in our FeSe samples the range of SC fluctuations, $\Delta T_{fl}$ (Table~I),was found to extend to temperatures exceeding double $T_c$, which is relatively much further above $T_c$ than is observed for cuprates.
The increased fluctuation contribution, $\Delta$ln$\sigma'$ (Table~I), was revealed in the range of 2D MT fluctuations, i.e. between $T_0$ and $T_{01}$.
Such an increase seems to be characteristic of magnetic superconductors \cite{SO,Sm} and this indicates a marked increase of magnetic interactions in FeSe as compared with non-magnetic YBCO. Nevertheless, as in cuprates, AL-MT (3D-2D) crossover at $T_0<T_{01}$ is observed for all our samples.
This allows us to determine a set of parameters which are important for fluctuation conductivity, such as the coherence length along the $c$-axis, $\xi_c(0) \simeq 3-5$~{\AA}, the distance between the Se atoms in the Se-Fe-Se conducting layer, $d_{01}$~$ \simeq 3$~{\AA}, and the phase relaxation time for fluctuating Cooper pairs, $\tau_{\phi}(25$K)~$ \simeq 3.4\cdot10^{-13}$~s.
It should be noted that for all samples, the values of $d_{01}$ (Table~I) are in good agreement with the results of structural studies.

Having determined the fluctuation conductivity parameters and using $R_H$(25K) we calculated a set of additional important properties of FeSe (Table~II) within a simple two-dimensional free-carrier model \cite{Mats,S3}. Eventually, rather unexpected results were obtained. Namely, the phase relaxation time turned out to be practically the same for FeSe samples having different values of other parameters, and even for YBCO films (Table~II).
This result, in turn, means that the range of SC fluctuations, $\Delta T_{fl}$, should be similar in different types of HTSCs. And indeed, for our samples,
$\Delta T_{fl}$ ranges from 9.6 K (sample S1) up to 20.5 K (sample S3) (Table~I) and, correspondingly, $\Delta T_{fl}$ is equal to 9.2~K for YBCO film (F1), and ranges from 7.0 to 16 K for YBCO-PrBCO superlattices and heterostructures studied previously \cite{We}.
Accordingly, $\Delta T_{fl} = 7$ and 23~K is reported in Ref. \cite{Grb}, for the nearly optimally doped ($T_c =$ 89 K) and deeply underdoped ($T_c =$ 57 K) YBCO single crystals, respectively. Moreover, similar temperature ranges of SC fluctuations have been found for FeAs-based compounds, $\Delta T_{fl} = 13.7$~K for SmFeAsO$_{0.85}$ \cite{Sm}, and $\Delta T_{fl} = 10$~K for EuFeAsO$_{0.85}$F$_{0.15}$ \cite{SO}.
All these results show that the range of SC fluctuations in many different types of HTSCs extends from about 5 to 25~K and that $\Delta T_{fl}$ is not proportional to $T_c$.

It is worth noting that sample S2, prepared by the solid state reaction method, has some parameters significantly different than the other two samples. Indeed, it has a much larger $\mu_H$, $l$, and $v_F$ and much smaller resistivity and effective mass of charge carriers, $m^*/m_0$ (Table~II). But, in spite of that and the fact that sample S2 has the smallest $\Delta T_{fl} = 6.6$~K, $\tau_{\phi}$(25K)~$=3.2\cdot10^{-13}$~s is determined in line with the results obtained for all other samples. This means that the phase relaxation time, which is the life time of the fluctuating Cooper pairs, is not the only parameter which determines the properties of SC fluctuations in FeSe and, additionally, that the properties strongly depend on the method of sample preparation.
Summing up the results presented in Table~I and Table~II, we can conclude that, somewhat unexpectedly, the Fermi velocity and rather large effective mass of charge carriers do not markedly affect the lifetime of the fluctuating Cooper pairs in the region of SC fluctuations between $T_0$ and $T_{01}$. However, the question what happens with the fluctuating Cooper pairs above $T_{01}$ still remains open.
The curves $\rho(T)$ (Fig.~1 and Refs. \cite{Kar,Kas,Ross}) and $R_H(T)$ \cite{Sun1,H2} do not show any characteristic features (anomalies) in the temperature range from $T_{01}$ to $T_S\approx 90$~K and this seems to imply that the local pairs in FeSe may exist at least up to $T_S$ \cite{Song,Pall,Ross}.
To test this assumption and to study the role of the possible local pairs formation at $T>>T_{01}$, additional extended research will need to be performed.\\

{\bf ACKNOWLEDGMENTS}

This work is partially conducted in the frame of Polish-Bulgarian and Polish-Ukrainian inter-academic research projects.\\

\end{document}